\documentclass[journal,10pt]{IEEEtran}
\usepackage[nolist]{acronym}
\usepackage{amsmath}

\usepackage{mathtools}
\usepackage[linesnumbered,ruled,vlined]{algorithm2e}
\usepackage{graphics}
\usepackage{breqn}

\usepackage{amssymb}
\usepackage{graphicx}

\graphicspath{{Figures/}}
\usepackage{xcolor} 

\usepackage{subcaption}

\begin{document}
\title{Enhancing Energy Efficiency in O-RAN Through Intelligent xApps Deployment}

\author{\IEEEauthorblockN{Xuanyu Liang\IEEEauthorrefmark{1}, Ahmed Al-Tahmeesschi\IEEEauthorrefmark{1}, Qiao Wang\IEEEauthorrefmark{1}, Swarna Chetty\IEEEauthorrefmark{1}, Chenrui Sun\IEEEauthorrefmark{1} and Hamed Ahmadi\IEEEauthorrefmark{1}}
\\
\IEEEauthorrefmark{1}School of Physics Engineering and Technology, University of York, United Kingdom\\
}

\maketitle

\begin{abstract}
The proliferation of 5G technology presents an unprecedented challenge in managing the energy consumption of densely deployed network infrastructures, particularly \acp{BS}, which account for the majority of power usage in mobile networks. The O-RAN architecture, with its emphasis on open and intelligent design, offers a promising framework to address the \ac{EE} demands of modern telecommunication systems. This paper introduces two xApps designed for the O-RAN architecture to optimize power savings without compromising the \ac{QoS}. Utilizing a commercial \ac{RIC} simulator, we demonstrate the effectiveness of our proposed xApps through extensive simulations that reflect real-world operational conditions. Our results show a significant reduction in power consumption, achieving up to 50\% power savings with a minimal number of \acp{UE}, by intelligently managing the operational state of \acp{RC}, particularly through switching between active and sleep modes based on network resource block usage conditions.
\end{abstract}

\begin{IEEEkeywords}
\ac{EE}, O-RAN, xApp, \ac{RIC}, \ac{QoS}
\end{IEEEkeywords}

\section{Introduction}
5G technology is poised to transform the industry with its high-speed data transmission and ultra-low latency, paving the way for a surge in new telecommunications services. However, this advancement necessitates a more densely packed network infrastructure, particularly in terms of access points, resulting in a proliferation of \acp{BS}. Consequently, energy consumption is likely to rise in correlation to the growing number of BS and the enhanced bit rate. Approximately 80\% of energy consumption in cellular mobile networks is attributed to \acp{BS} \cite{lahdekorpi2017energy}. The surge in energy consumption not only amplifies the operational costs for wireless network providers but also contributes to a rise in CO2 emissions \cite{auer2011much}. In this context, the O-RAN Alliance offer a new architecture and standard to achieving the various service requirements while enhancing Energy Efficiency (EE), which is one of fundamental \ac{KPIs} of growing importance as telecommunications companies face increasing operational costs \cite{pamuklu2020grove}, characterized by its open and intelligent design, facilitates a more efficient and versatile network infrastructure through the disaggregation of user and control plane functions and the integration of open interfaces and standards \cite{O-RanArchitecture, O-RanA1Interface}.

\ac{O-RAN} revolutionizes the conventional monolithic structure by disaggregating base station functionalities into distinct components: the O-\ac{CU}, the O-\ac{DU}, and the O-\ac{RU}. In this architecture, the O-\ac{CU} and O-\ac{DU} collectively embody the functionalities of the \ac{BBU} found in previous RAN models, with their operations predominantly hosted in the cloud. The O-\ac{RU}, on the other hand, serves as a physical entity, akin to the \ac{RRH} of traditional systems, albeit with a streamlined role. It is specifically tasked with managing the low-physical layer, the RF module and power amplifiers, which is a major contributor of the total energy consumption of \ac{O-RAN} \cite{abubakar2023energy}. Consequently, achieving a higher number of deactivated O-\acp{RU} directly correlates with increased power savings, provided that such reductions align with the load requirements of the O-\ac{RU} and the overall network demand.

\subsection{Key Innovations of O-RAN}
\begin{figure*}[!htbp]\label{workflow}
\vspace{-2mm}
	\centering
	\includegraphics[clip, trim=0.0cm 0.cm 0.0cm 0cm, width=2\columnwidth]{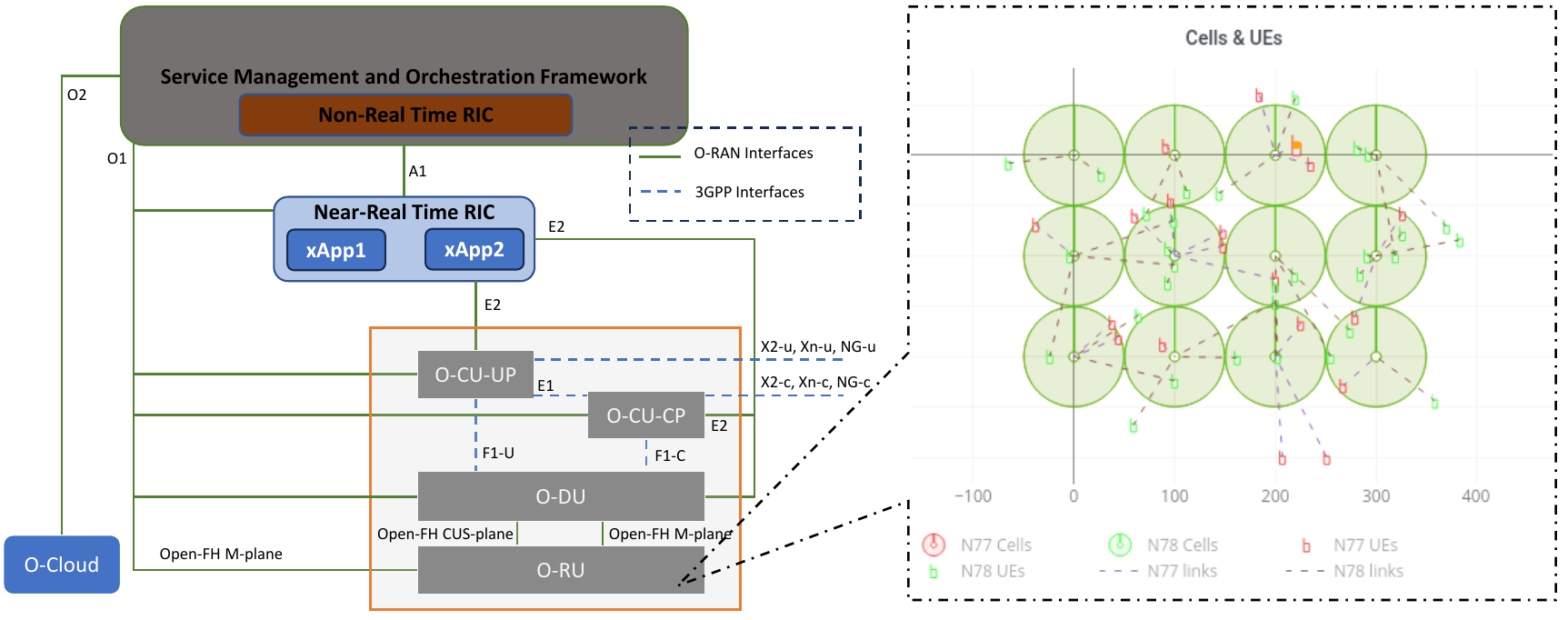}
	\caption{shows O-RAN Architecture, with components and interfaces from O-RAN and 3GPP. O-RAN interfaces are drawn as solid lines, 3GPP ones as dashed lines. The layout of the O-RU and the distribution of users is shown in the picture in the right section. Each O-RU consists of two \acp{RC} operating on distinct frequencies, n77 and n78, represented by red and green circles, respectively. Due to the identical coverage areas of both \acp{RC}, only the color of one RC is discernible in the figure. The color of the \acp{UE} within the figure matches the RC to which they are connected, indicating their network association based on the respective \ac{RC}'s frequency. The proposed two xApps are anchored in Near-RT RIC}
	\label{fig:systemmodel}
\end{figure*}

A significant innovation within the O-RAN framework is the introduction of the \ac{RIC} illustrated in Fig. \ref{fig:systemmodel}, a programmable unit crafted to manage the growing complexities of network infrastructures effectively, thus improving both the precision and the operational performance. Acting as a pivotal abstraction layer within the network, the RIC aggregates extensive \ac{KPMs} data, offering a detailed overview of the network infrastructure's condition, including user metrics, traffic loads, and throughput capabilities \cite{pamuklu2021energy}. To cater to various operational dynamics, the architecture includes two distinct versions of the RIC: the \ac{Near-RT RIC} and the \ac{Non-RT RIC}  \cite{Non-RTRICfunctionalarchitecture} which anchored in orchestrating network management, each tasked across different temporal scales, specifically within time frames of 10 ms to 1s, and beyond 1s, in a closed-loop system. The \ac{Near-RT RIC} consists multiple applications to support various service demands called xApps. xApps are developed using a standardized framework that allows for their deployment across different vendor equipment and RAN configurations \cite{O-RanNear-RTRIC2.00} via the open interfaces. Within the framework established by the O-RAN Alliance, open interfaces are precisely defined standard interfaces that foster a diverse ecosystem of options for operators. These interfaces enable operators to select from an array of vendors for various O-RAN components, thereby enhancing market competitiveness. Among these, the E2 interface stands out as a critical conduit linking the \ac{Near-RT RIC} with E2 nodes \cite{E2InterfaceTest}, which include \ac{DU}, \ac{CU}, and LTE eNB compliant with O-RAN specifications. The principal role of the E2 interface revolves around the acquisition of dynamic network information from E2 nodes, with the Near-RT RIC subsequently processing this data and generating actionable feedback. In our case, feedback from Near-RT RIC can be the set of active \acp{RC}. Another pivotal open interface within the O-RAN architecture is the O1 interface. This interface is instrumental in connecting all O-RAN elements, encompassing RAN nodes and the Near-RT RIC, with the \ac{SMO} framework and the Non-Real Time RIC \cite{O1InterfaceTest}. The O1 interface significantly empowers the SMO, which is tasked with the comprehensive management, orchestration, and automation of network services and functionalities, enabling direct access to network information.

\subsection{Related Works and Contributions}

Some studies focus on using mathematical optimization to select the proper switching-off \ac{RRH} or \ac{BS} without considering the user's \ac{QoS} \cite{feng2017boost, alemam2019energy}. Those approaches show a decent performance on energy efficiency, but it takes high computational requirements and increases the network complexity when the network is laid out densely. 

The authors in \cite{oh2016unified} mainly discussed the dynamic operation of macro \acp{BS} including both switch off and on for potential energy saving. Both \ac{DL} and \ac{UL} were taken into consideration. The proposed algorithm allowed for a flexible balance between \ac{UL} and \ac{DL} by adjusting a parameter called the fraction factor. However, the authors did not study the impact of energy saving on \ac{UE}'s \ac{QoS}.

Authors in \cite{oh2013dynamic} and \cite{jian2019energy} formulated the \ac{BS} switching strategies as a joint optimization problem, taking user's \ac{QoS} and user association policy into consideration. The algorithm \cite{oh2013dynamic}  aimed to turn off \acp{BS} one by one, taking into account the additional load increments brought to neighboring \acp{BS}. The paper also proposed three heuristic versions of \ac{SWES} that use approximate values of network-impact as decision metrics to reduce signaling and implementation overhead further. In \cite{jian2019energy}, the authors tackled the \ac{EEMP} in two phases. First, it examines static-\ac{EEMP}, using heuristic algorithms to optimize energy efficiency by considering user proximity and base station impacts. Then, it introduces the \ac{QFMEE} strategy, which iteratively deactivates base stations, reassigns users, and evaluates short-term network efficiency, seeking an optimal network configuration that maintains service quality. While this paper is the user-centric putting \ac{QoS} in prime position, the process is overly complex and it is possible to integrate two steps into a single step.

On the other hand, \cite{wu2021deep, rezaei2023energy, masoudi2020reinforcement, liu2018deepnap} control the \ac{BS} switching off or \ac{BS} sleep mode by adapting \ac{DRL}. The researchers segmented the sleep mode of the \ac{BS} into various stages, each characterized by distinct levels of energy consumption and specific conditions for activation, which significantly enhanced the optimization of energy efficiency across the entire network. \ac{ML} models typically demand greater computational resources relative to certain mathematical methodologies. Additionally, they necessitate substantial volumes of data for training purposes, along with extended periods dedicated to the training process.

The purpose of this paper is to create an energy saving xApp and implement it into a realistic environment. In addition, our work differ from previous literature in the assumption of having multiple \acp{RC} in an O-\ac{RU}. An \ac{RC} within an O-\ac{RU} is a hardware module responsible for converting digital signals to radio frequencies and vice versa. It carries out key operations like modulation, demodulation, amplification, and filtering of signals. Each RC operates on specific frequency bands, allowing the O-RU to support various cellular standards and ensure effective communication between the network and \acp{UE}. We propose two xApps to control the sleep mode of \acp{RC} while satisfying the UE's \ac{QoS} and load threshold of \acp{RC}. Both xApps are implemented into the real simulator `TeraVM' RIC-tester \cite{rictester} and obtain a maximum of 50\% power saving during small number of \acp{UE}. 

The main contributions of this paper are:
\begin{itemize}
    \item We develop two xApps for \acp{RC} switching within the O-RAN architecture to reduce the network's overall power consumption.
    \item We integrate those xApps with the 'TeraVM' RIC-tester, a real simulation software environment, demonstrating its practical applicability.
    \item We justify the achievable performance of the proposed approach in terms of power saving with simulation and demonstrate the effectiveness of the xApps in adapting to changes in the radio environment.
\end{itemize}

\section{System Model and Environment}
In this segment, the configuration of the system implemented in this study is delineated. To craft a simulation environment that mirrors real-world conditions closely, the 'TeraVM' \ac{RIC} tester is employed. Additionally, the \ac{UMa} channel model is utilized to enhance the realism of the simulations \cite{rictester}. This research incorporates a single O-\ac{CU} and 12 O-RUs, each of which is serviced by a corresponding O-\ac{DU}. The O-\acp{RU} are positioned at fixed locations, maintaining a uniform inter-site distance of 100 meters. The installation height for each O-RU is standardized at 10 meters, and each O-RU encompasses two \acp{RC} operating on distinct frequencies—specifically, n77 and n78 as illustrated in Fig. \ref{fig:systemmodel}. Both \acp{RC} are equipped with isotropic antennas, and the transmission power for each RC is set to 30 dBm. During the simulation, all the \acp{UE} are static and randomly distributed in the considered area. For the sake of subsequent better expression, we assume that there are a set of \acp{RC} $\textbf{M}$ and a set of \acp{UE} $\textbf{K}$. Let $k \in \textbf{K}$ be the index of $k$-th \ac{UE} and $m \in \textbf{M}$ be the index of $m$-th \ac{RC}. The power consumption model for \acp{RC} are delineated following the framework presented by \cite{bjornson2015optimal}. The model encapsulates the various components of power expenditure as follows:
\begin{equation}
    P_{RC} = P_{FIX}^{mode}+P_{data}+P_{TC},
\end{equation}
where, $P_{RC}$ represents the overall power consumption of all \acp{RC} in the network. $P_{TC}$ denotes for the power consumed by the transceiver chains. The term $P_{FIX}^{mode}$ represents the fixed power consumption of \acp{RC}, which varies depending on whether an RC is in its active state or sleep mode. This can be expressed as:
\begin{equation}
    P_{FIX}^{mode} = \sum_{m = 1}^{M}\alpha_m*P_{Active}+(1-\alpha_m)*P_{Sleep},
\end{equation}
in which $\alpha_m$ is a binary  indicator denoting the operational state of RC $m$ ($1$ for active and $0$ for sleep). The $P_{Active}$ and $P_{Sleep}$ respectively signify the fixed power consumption rates of an RC in its active and sleep states. Furthermore, accounts for the power consumption associated with the transmission power of RC $m$, which scales with the usage of RC resource blocks as defined by:
\begin{equation}
    P_{data} = \sum_{m = 1}^{M}\alpha_m*\frac{P_{tx}}{\eta }* PRB_{usage}^{m},
\end{equation}
where the $P_{tx}$ is the transmission power of \ac{RC} and $\eta$ is represented as the \ac{PA} efficiency. The term $PRB_{usage}^{m}$ reflects the resource block usage on RC $m$, illustrating that an increase in the number of \acp{UE} served by an RC directly influences its power consumption.

\section{Problem Formulation}

The objective function of this study is to minimize the total power consumption of the network. To achieve this, we have developed two xApps that adjusts the activity of \acp{RC} based on current network conditions. The xApps aim to find the most energy-efficient configuration of an \acp{RC} without compromising network performance and quality. This optimization is subject to three key constraints. The first is \ac{RSRP}, which represents the power of the reference signals spread over the entire bandwidth and is measured in decibels milliwatt (dBm). \ac{RSRP} is a vital indicator of signal strength received by a UE from an RC and plays a significant role in determining the quality of the connection between the \ac{UE} and the network. By ensuring that each \ac{UE} maintains a minimum \ac{UE} level, we aim to guarantee robust and reliable connectivity across the network.

In addition, data rate as the second constrain ensures that the data rate meets the operational needs of the \acp{UE}. The last constrain is managing resource block usage on each RC efficiently. This resource block management prevents any RC from becoming overwhelmed with traffic, ensuring a balanced distribution of network resources.
Consequently, the resource block utilization $L_m$ of RC $m$ is defined as the average portion of bandwidth occupied:
\begin{equation}
    L_m = \sum_{k\in\textbf{K}_m}B_k/B_m,
\end{equation}
where $B_m$ denotes the maximum resource block of \ac{RC} $m$ and $B_k$ represents the resource blocks currently occupied by the \ac{UE} $k$.
Hence, the problem of minimizing the network's power consumption is formulated as:
\begin{equation}
\label{objective function}
\begin{aligned}
\mathcal{P}_1: \quad \text{min} \quad &  P_{RC} \\
\text{s.t.} \quad & \sum_{m =1}^{M} \beta_{k,m}\gamma _{k,m}\geq \gamma_{min}, \quad \forall k \in \textbf{K}, \\
                    & \sum_{m =1}^{M} \beta_{k,m} R_{k,m}\geq R_{min}, \quad \forall k \in \textbf{K},\\
                    & \sum_{m =1}^{M}\beta_{k,m}\leq 1 \quad \forall k \in \textbf{K},\\
                    & L_m\leq 1, \quad \forall m \in \textbf{M},
\end{aligned}
\end{equation}
where \(\beta_{k,m}\) is the binary variable to represent the association between the \ac{UE} $k$ and \ac{RC} $m$. $\beta_{k,m} = 1$ means \ac{UE} $k$ and \ac{RC} $m$ is associated. \(\gamma_{min}\) is the minimum \ac{RSRP} threshold necessary for \ac{UE} $k$ and $\gamma_{k,m}$ is the \ac{RSRP} of \ac{UE} $k$ from \ac{RC} $m$. $R_{k,m}$ is the data rate of \ac{UE} $k$ when associate with RC $m$ and $R_{min}$ is the minimum required data rate by a \ac{UE}. A \ac{UE} is considered to be in an outage if its data rate is below this specified threshold. Additionally, to prevent any RC from being overloaded, $L_m$ must not exceed 1.

\section{\ac{RC} Switching Algorithm}
The objective, as stated in \eqref{objective function}, is to identify an optimal set of active \acp{RC} that minimises power consumption. To achieve this goal, we propose two distinct xApps. One of the clear benefits of the considered approach is that all the \ac{KPMs} are all defined by 3GPP and O-RAN Alliance.

\subsection{Proposed xApp1}
Given that \acp{UE} are randomly distributed within a designated area, this results in not all \acp{RC} being associated with \acp{UE}. Especially when the number of \acp{UE} is relatively small, there will be some \acp{RC} in idle state (i.e., not serving \acp{UE}). Our designed first xApp is to switch those \acp{RC} in to sleep mode, which illustrate on Algorithm 1. First of all it iterates through all the RCs in the RC list and then utilizes the KPM of RRC.ConnMean to assess the number of \acp{UE} serviced by each \ac{RC} \cite{RRC}. If an \ac{RC} is found to have no \acp{UE} to serve. The xApp to transition that \ac{RC} into sleep mode. 

\begin{algorithm}
\caption{xApp1 algorithm}
\DontPrintSemicolon  
\For{each RC in \textbf{M}}{
    Retrieve the list of \acp{UE} attached to the current RC\;
    \If{the list of attached \acp{UE} is empty}{
        Put the current RC into sleep mode\;
    }
}
\end{algorithm}

\subsection{Proposed xApp2}
The Algorithm 2 is proposed based on the initial intervention by the first xApp, certain \acp{RC} are set to sleep mode to enhance network efficiency. This action, however, leads to a new scenario where some of the still-active \acp{RC} are found to be serving a very small number of \acp{UE}, indicating a low-load operation. These \acp{RC}, due to their minimal usage, become prime candidates for potentially entering sleep mode to conserve power. In the proposed algorithm, when both of resource block usage and throughput below the certain threshold can be regarded as low-load \acp{RC}.

The possibility of transitioning these low-load \acp{RC} into sleep mode hinges on successfully reallocating the \acp{UE} they serve to other RCs. This reallocation must meet two critical conditions: firstly, the \ac{RSRP} for the \acp{UE} must remain above a certain threshold to ensure no compromise on service quality; secondly, the load on any new \ac{RC} tasked with serving additional \acp{UE} should not exceed a specific limit, for instance, 50\%, to prevent overburdening. When all the \acp{UE} in the RC have been successfully reassigned out, the RC can enter the sleep state. The reassigned \ac{UE} needs to update its RSS based on the new association. But when the \ac{UE} cannot fulfil both hard constrains. The RC must remain active even if it is in a low load state.

The proposed algorithm seeks to further diminish the network's total power usage. This ensures that the network capacity is utilised appropriately without unnecessarily keeping under-loaded \acp{RC} active. This xApp presents a balanced way to optimize network performance while minimizing power expenditure.

\begin{algorithm}
\caption{xApp2 algorithm}
\DontPrintSemicolon
Initialize Algorithm 1\;
\For{each active RC $m$}{
    \If{the resource block usage and throughput of RC $m$ less than threshold}{
        Select those \acp{RC}\;
        \For{each UE $k$ served by selected RC $m$}{
            \For{each RC $n(n\in\textbf{M},n\neq m)$ in the network}{
                \uIf{the RSRP of UE $k$ is satisfied threshold $\gamma_{min}$ and load of RC $L_{n}$ can accept extra UE}{
                    Reassign the UE to RC $n$ and update the UE's RSRP based on the new association.\;
                }
                \Else{
                    RC $m$ can't be switched into sleep mode\;
                }
            }
        }
        \If{all the UE $k$ served by RC $m$ are re-associated}{
            Switch RC $m$ into sleep mode\;
        }
    }
}
\end{algorithm}

\begin{figure}[!t]
	\centering
	\includegraphics[clip, trim=0.0cm 0.cm 0.0cm 0cm, width=1\columnwidth]{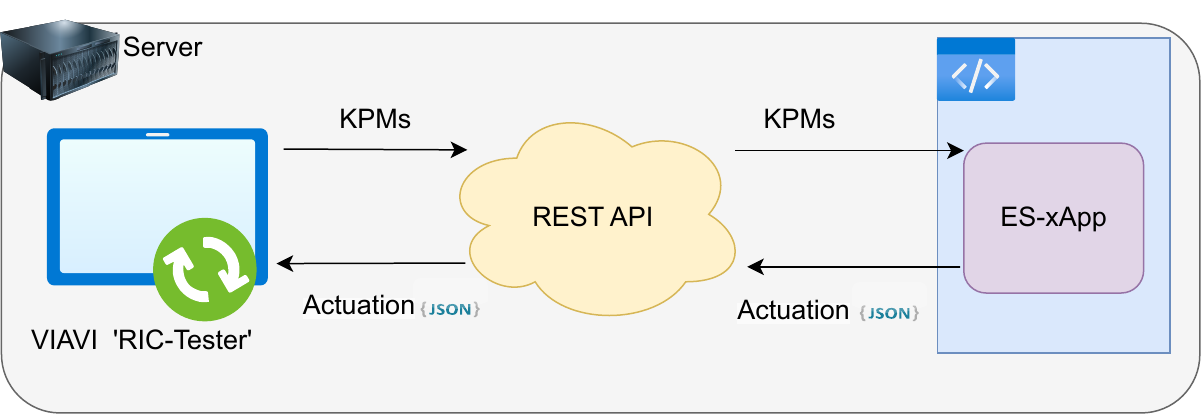}
	\caption{shows the flowgraph that how the python code based xApp interact with RIC-tester.}
    \label{workflow1}
\end{figure}

\section{Results}

We consider a network consisting of 12 O-\acp{RU} and 24 \acp{RC} in the area of 500 $\times$ 500 m$^2$, where the inter-site-distance is 100m. We randomly distributed 10, 50 and 100 \acp{UE}. The channel model is chosen to follow the \ac{UMa} model \cite{Uma}.

\begin{figure}[!t]\label{fig:Mean_avg_PC}
	\centering
	\includegraphics[clip, trim=0.0cm 0.cm 0.0cm 0cm, width=1\columnwidth]{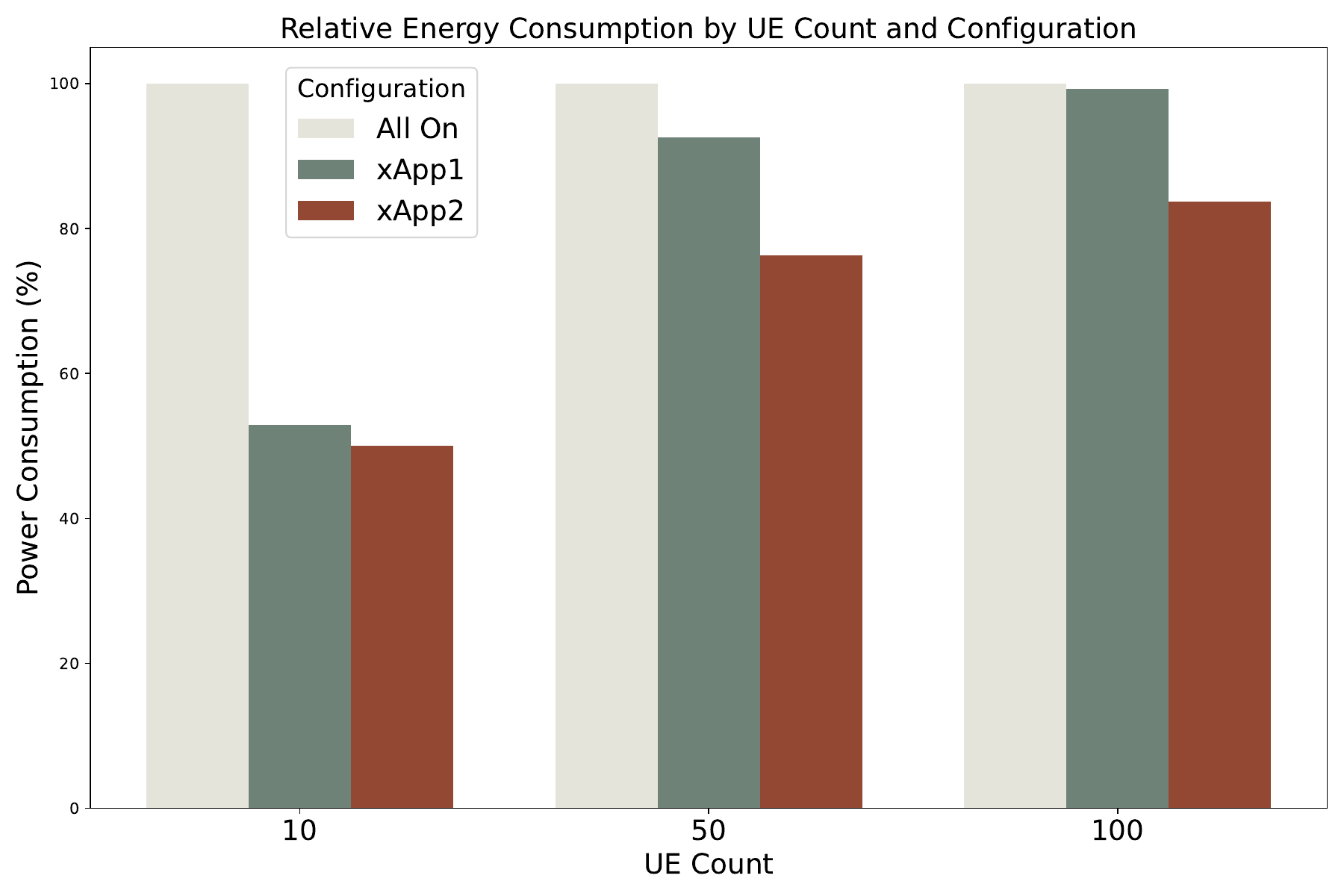}
	\caption{shows the power consumption in percentage among 10, 50 and 100 \acp{UE} in three different scenarios. All on is always in 100\% power consumption. xApp1 and xApp2 show a relative percentage of the All on scenario. }
    \label{fig:Mean_avg_PC}
\end{figure}

The entire procedure is systematically depicted in Fig. \ref{workflow1}. Initially, an xApp initiates the process by dispatching a request for network data or \ac{KPMs} to the \ac{RIC}-tester through the REST \ac{API}. Subsequently, the \ac{RIC}-tester processes this request and furnishes the requisite feedback to the xApp. Upon receiving the feedback, the xApp evaluates the appropriate action based on its embedded algorithm and communicates the decision back to the RIC-tester using the 'REST' API. Finally, the \ac{RIC}-tester implements the specified configuration in accordance with the action determined by the xApp.

Our initial exploration focuses on assessing the average (over 10 trials) power consumption in percentage across varying distributions of UE utilizing our proposed algorithms. Illustrated in Fig. \ref{fig:Mean_avg_PC}, with a scenario of merely 10 \acp{UE}, both the xApp1 and the xApp2 demonstrate substantial power savings, reaching up to 50\%. In this context, the performance differential between the xApp1 and xApp2 is less than 5\% because with a small number of \acp{UE}, xApp1 has already switched most of the RCs into sleep mode, and xApp2 can no longer find RCs that fulfil the conditions. However, as the UE count escalates, a divergence in performance becomes apparent. Specifically, when the UE tally extends to 100, the xApp1's power consumption parallels that observed when all RCs are active, contrasting sharply with xApp2, which sustains approximately 15\% in power savings. On top of that, xApp2 still outperforms xApp1 by almost 20\%. This analysis underscores the efficiency of xApp2 in managing power consumption, particularly as network demand intensifies.

\begin{figure}[!t]
	\centering
	\includegraphics[clip, trim=0.0cm 0.cm 0.0cm 0cm, width=1\columnwidth]{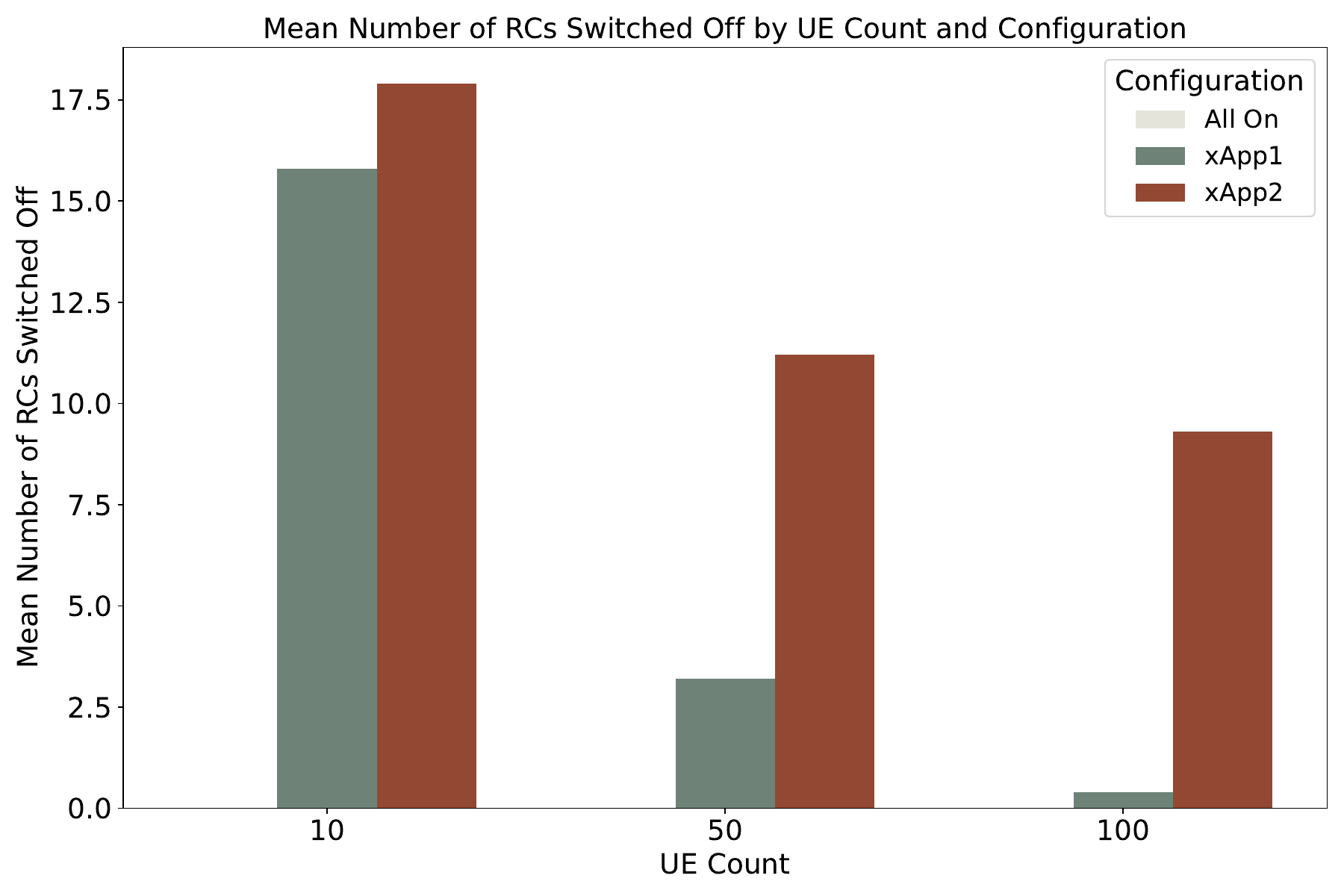}
	\caption{shows the average (over 10 trials) number of RCs are switched into sleep mode between 10, 50 and 100 UEs in three different scenarios}
    \label{fig:Mean_avg_RC_off}
\end{figure}

In Fig. \ref{fig:Mean_avg_RC_off} we examine the number of RCs are switched into sleep mode. It becomes evident that under conditions of high \ac{UE} density, the xApp1 exhibits a limited propensity to transition RCs into sleep mode. This observation is corresponding to the data presented in Fig. \ref{fig:Mean_avg_PC}. Conversely, the xApp2, leveraging our proposed algorithm, demonstrates the capability to place approximately 10 RCs into sleep mode, even with increased \ac{UE} density (50 and 100). A comparative analysis of Fig. \ref{fig:Mean_avg_PC} and Fig. \ref{fig:Mean_avg_RC_off} reveals that, for xApp2, the variation in the number of RCs entering sleep mode between scenarios with 50 and 100 UEs is marginal, at around 2 RCs. Nonetheless, this small difference leads to a more noticeable difference in power savings. This phenomenon can be attributed to the escalating demands placed on RCs as the number of UEs increased. Not only are RCs required to accommodate a larger volume of UEs, but they must also manage UEs located at greater distances. This scenario necessitates increased resource block usage utilization and, consequently, leads to heightened power consumption.

\begin{table}[]
\centering
\caption{Simulation Parameters}
\begin{tabular}{|l|l|}
\hline
Parameter                                        & Value                   \\ \hline
Carrier frequency $n77/ n78$                     & 3.5/3.7GH                         \\ \hline
Channel bandwidth, $B_m$                          & 100MHz                   \\ \hline
Fixed power in active mode, $P_{Active}$                           & 20W                   \\ \hline
Fixed power in sleep mode, $P_{Sleep}$                   & 5W                     \\ \hline
Power Amplifier efficiency (PA), $\eta$                 & 1.67\%                  \\ \hline
Power per RF chain, $P_{TC}$                            & 1W                    \\ \hline
UE required data rate, $R_{min}$        & 10 Mbps                     \\ \hline
The number of O-RUs, $\textbf{N}$        & 12                     \\ \hline
The number of RCs, $\textbf{M}$        & 24                      \\ \hline

\end{tabular}
\label{table2}
\end{table}

\section{Conclusions}

In this paper, we introduced two xApps specifically designed for switching \acp{RC} into sleep mode while satisfying \ac{UE} required data rate and resource block usage of \acp{RC}. This work also tested within the ``TeraVM'' RIC-tester, a commercial-grade simulation environment. Our numerical results substantiate the effectiveness of the proposed xApps in achieving considerable power savings across varying network demands. Moving forward, our focus will include considering user mobility, thereby treating traffic variations in O-RUs as an ongoing dynamic process. Machine learning techniques, such as deep reinforcement learning, are expected to offer superior performance in navigating the complexities and constraints of network environments.

\section*{Acknowledgement}

This work was supported in part by the Engineering and Physical Sciences Research Council United Kingdom (EPSRC), Impact
Acceleration Accounts (IAA) (Green Secure and Privacy Aware Wireless Networks for Sustainable Future Connected and Autonomous
Systems) under Grant EP/X525856/1 and Department of Science, Innovation and Technology, United Kingdom, under Grants Yorkshire
Open-RAN (YO-RAN) TS/X013758/1 and RIC Enabled (CF-)mMIMO for HDD (REACH) TS/Y008952/1.

\begin{acronym} 
\acro{5G}{Fifth Generation}
\acro{ACO}{Ant Colony Optimization}
\acro{ANN}{Artificial Neural Network}
\acro{BB}{Base Band}
\acro{BBU}{Base Band Unit}
\acro{BER}{Bit Error Rate}
\acro{BS}{Base Station}
\acro{BSs}{Base Stations}
\acro{BW}{bandwidth}
\acro{C-RAN}{Cloud Radio Access Networks}
\acro{CAPEX}{Capital Expenditure}
\acro{CoMP}{Coordinated Multipoint}
\acro{CR}{Cognitive Radio}
\acro{D2D}{Device-to-Device}
\acro{DAC}{Digital-to-Analog Converter}
\acro{DAS}{Distributed Antenna Systems}
\acro{DBA}{Dynamic Bandwidth Allocation}
\acro{DC}{Duty Cycle}
\acro{DL}{Deep Learning}
\acro{DSA}{Dynamic Spectrum Access}
\acro{FBMC}{Filterbank Multicarrier}
\acro{FEC}{Forward Error Correction}
\acro{FFR}{Fractional Frequency Reuse}
\acro{FSO}{Free Space Optics}
\acro{GA}{Genetic Algorithms}
\acro{HAP}{High Altitude Platform}
\acro{HL}{Higher Layer}
\acro{HARQ}{Hybrid-Automatic Repeat Request}
\acro{HCA}{Hierarchical Cluster Analysis}
\acro{HO}{Handover}
\acro{KNN}{k-nearest neighbors} 
\acro{IoT}{Internet of Things}
\acro{LAN}{Local Area Network}
\acro{LAP}{Low Altitude Platform}
\acro{LL}{Lower Layer}
\acro{LoS}{Line of Sight}
\acro{LTE}{Long Term Evolution}
\acro{LTE-A}{Long Term Evolution Advanced}
\acro{MAC}{Medium Access Control}
\acro{MAP}{Medium Altitude Platform}
\acro{MDP}{Markov Decision Process}
\acro{ML}{Machine Learning}
\acro{MME}{Mobility Management Entity}
\acro{mmWave}{millimeter Wave}
\acro{MIMO}{Multiple Input Multiple Output}
\acro{NFP}{Network Flying Platform}
\acro{NFPs}{Network Flying Platforms}
\acro{NLoS}{Non-Line of Sight}
\acro{OFDM}{Orthogonal Frequency Division Multiplexing}
\acro{O-RAN}{Open Radio Access Network}
\acro{OSA}{Opportunistic Spectrum Access}
\acro{PAM}{Pulse Amplitude Modulation}
\acro{PAPR}{Peak-to-Average Power Ratio}
\acro{PGW}{Packet Gateway}
\acro{PHY}{physical layer}
\acro{PSO}{Particle Swarm Optimization}
\acro{PU}{Primary User}
\acro{QAM}{Quadrature Amplitude Modulation}
\acro{QoE}{Quality of Experience}
\acro{QoS}{Quality of Service}
\acro{QPSK}{Quadrature Phase Shift Keying}
\acro{RF}{Radio Frequency}
\acro{RL}{Reinforcement Learning}
\acro{RMSE}{Root Mean Squared Error}
\acro{RN}{Remote Node}
\acro{RRH}{Remote Radio Head}
\acro{RRC}{Radio Resource Control}
\acro{RRU}{Remote Radio Unit}
\acro{RSS}{Received Signal Strength}
\acro{SU}{Secondary User}
\acro{SCBS}{Small Cell Base Station}
\acro{SDN}{Software Defined Network}
\acro{SNR}{Signal-to-Noise Ratio}
\acro{SON}{Self-organising Network}
\acro{SVM}{Support Vector Machine}
\acro{TDD}{Time Division Duplex}
\acro{TD-LTE}{Time Division LTE}
\acro{TDM}{Time Division Multiplexing}
\acro{TDMA}{Time Division Multiple Access}
\acro{UE}{User Equipment}
\acro{UAV}{Unmanned Aerial Vehicle}
\acro{USRP}{Universal Software Radio Platform}
\acro{DRL}{Deep Reinforcement Learning}
\acro{AI}{Artificial Intelligence}
\acro{RAN}{Radio Access Network}
\acro{RU}{Radio Unit}
\acro{CU}{Central Unit}
\acro{DU}{Distributed Unit}
\acro{NR}{New Radio}
\acro{gNBs}{Next Generation Node Bases}
\acro{CP}{Control Plane}
\acro{UP}{User Plane}
\acro{FPGAs}{Field  Programmable  Gate  Arrays}
\acro{ASICs}{Application-specific Integrated Circuits}
\acro{PHY-low}{lower level PHY layer processing}
\acro{FFT}{Fast Fourier Transform}
\acro{RRC}{Radio Resource Control}
\acro{SDAP}{Service Data Adaptation Protocol}
\acro{PDCP}{Packet Data Convergence Protocol}
\acro{RLC}{Radio Link Control}
\acro{RIC}{RAN Intelligent Controller}
\acro{RICs}{RAN Intelligent Controllers}
\acro{KPMs}{Key Performence Measurements}
\acro{RT}{Real Time}
\acro{SMO}{Service Management and Orchestration}
\acro{UE}{User Equipment}
\acro{API}{Application Programming Interface}
\acro{OSC}{O-RAN Software Community}
\acro{DRL}{Deep Reinforcement Learning}
\acro{OSP}{Online Service Provider}
\acro{NIB}{Network Information Base}
\acro{SDL}{Shared Data Layer}
\acro{SLA}{Service Level Agreement}
\acro{A1AP}{A1 Application Protocol}
\acro{HTTP}{Hypertext Transfer Protocol}
\acro{SL}{Supervised Learning}
\acro{UL}{Unsupervised Learning}
\acro{RL}{Reinforcement Learning}
\acro{DL}{Deep Learning}
\acro{FDD}{Frequency-division Duple}
\acro{TDD}{Time-division Duple}
\acro{LSTM}{Long Short-term Memory}
\acro{PCA}{Principal Component Analysis}
\acro{ICA}{Independent Component Analysis}
\acro{MDP}{Markov Decision Process}
\acro{GRL}{Generalization Representation Learning}
\acro{SRL}{Specialization Representation Learning}
\acro{SVM}{Support Vector Machine}
\acro{TDNN}{Time-delay Neural Network}
\acro{LSTM}{Long Short-term Memory}
\acro{MSE}{Mean Squared Error}
\acro{CNN}{Conventional Neural Network}
\acro{NAS}{Neural Architecture Search}
\acro{SDS}{Software Defined Security}
\acro{SON}{Self-organized Network}
\acro{KPIs}{Key Performance Indicators}
\acro{HetNet}{Heterogeneous Network}
\acro{HPN}{High Power Node}
\acro{LPNs}{Low Power Nodes}
\acro{QL}{Q-Learning}
\acro{PG}{Policy Gradient}
\acro{A2C}{Actor-Critic}
\acro{TD}{Temporal Defence}
\acro{SHAP}{SHapley Additive exPlanations}
\acro{DNNs}{Deep Neural Networks}
\acro{DNN}{Deep Neural Network}
\acro{MLP}{Multiple-layer Perceptron}
\acro{RNN}{Recurrent Neural Network}
\acro{DRL}{Deep Reinforcement Learning}
\acro{DQN}{Deep Q-Learning Network}
\acro{DDQN}{Double Deep Q-Learning Network}
\acro{GNN}{Graph Neural Network}
\acro{eMMB}{enhanced mobile broadband}
\acro{URLLC}{ultra-reliable low-latency communication}
\acro{QoS}{Quality of Service}
\acro{ILP}{Integer Linear Programming}
\acro{NF}{Network Function}
\acro{VFN}{Network Function Virtualization}
\acro{NBC}{Navie Bayes Classifier}
\acro{RAT}{Radio Access Technology}
\acro{FML}{federated meta-learning}
\acro{TS}{Traffic Steering}
\acro{CQL}{Conservative Q-Learning}
\acro{REM}{Random Ensemble Mixture}
\acro{SCA}{Successive Convex Approximation}
\acro{XAI}{eXplainable Artificial Intelligent}
\acro{D-RAN}{Distributed RAN}
\acro{C-RAN}{Cloud RAN}
\acro{v-RAN}{Virtual RAN}
\acro{RRH}{Remote Radio Head}
\acro{mMTC}{massive machine-type communication}
\acro{CDMA}{Code Division Multiple Access}
\acro{TDMA}{Time Division Multiple Access}
\acro{OFDMA}{Orthogonal Frequency-Division Multiple Access}
\acro{RRM}{Radio Resources Management}
\acro{NFV}{Network Function Virtualization}
\acro{D-RAN}{Distributed-RAN}
\acro{V-RAN}{Vritualized-RAN}
\acro{PA}{Power Amplifier}
\acro{SINR}{Signal to interference plus noise ratio}
\acro{mmWave}{millimeter Wave }
\acro{LOS}{Line of Sight}
\acro{NLOS}{Non-Line of Sight}
\acro{FSPL}{Free Space Path Loss}
\acro{EEMP}{Energy Efficiency Maximization Problem}
\acro{QFMEE}{QoS First Maximum EE}
\acro{SWES}{Switching-on/off based Energy Saving}
\acro{APC}{Area Power Consumption}
\acro{EE}{Energy Efficiency}
\acro{Near-RT}{Near-Real Time}
\acro{Open-RAN}{Open Radio Access Network}
\acro{Near-RT RIC}{Near Real Time RIC}
\acro{Non-RT RIC}{Non Real Time RIC}

\acro{RC}{Radio Card}
\acro{UMa}{Urban Microcell path loss}

\acro{API}{Application Programming Interface}
\acro{RSS}{Received Signal Strength}
\acro{DL}{downlink}
\acro{UL}{uplink}
\acro{RSRP}{Reference Signals Received Power}
\end{acronym}

\bibliographystyle{IEEEtran}

\bibliography{References.bib}
\end{document}